\documentclass[preprint]{aastex}

\begin{document}

\newcommand{\lya}{Lyman-$\alpha$}
\newcommand{\eqw}{\hbox{EW}}
\def\erg{\hbox{erg}}
\def\cm{\hbox{cm}}
\def\sec{\hbox{s}}
\def\f17{f_{17}}
\def\Mpc{\hbox{Mpc}}
\def\Gpc{\hbox{Gpc}}
\def\nm{\hbox{nm}}
\def\km{\hbox{km}}
\def\kms{\hbox{km s$^{-1}$}}
\def\year{\hbox{yr}}
\def\Myr{\hbox{Myr}}
\def\Gyr{\hbox{Gyr}}
\def\deg{\hbox{deg}}
\def\arcsec{\hbox{arcsec}}
\def\microJy{\mu\hbox{Jy}}
\def\zre{z_r}
\def\fesc{f_{\rm esc}}

\def\ergcm2s{\ifmmode {\rm\,erg\,cm^{-2}\,s^{-1}}\else
                ${\rm\,ergs\,cm^{-2}\,s^{-1}}$\fi}
\def\ergsec{\ifmmode {\rm\,erg\,s^{-1}}\else
                ${\rm\,ergs\,s^{-1}}$\fi}
\def\kmsMpc{\ifmmode {\rm\,km\,s^{-1}\,Mpc^{-1}}\else
                ${\rm\,km\,s^{-1}\,Mpc^{-1}}$\fi}
\def\kpc{{\rm kpc}}
\def\nv{\ion{N}{5} $\lambda$1240}
\def\civ{\ion{C}{4} $\lambda$1549}
\def\oii{[\ion{O}{2}] $\lambda$3727}
\def\oiipair{[\ion{O}{2}] $\lambda \lambda$3726,3729}
\def\oiii{[\ion{O}{3}] $\lambda$5007}
\def\oiiib{[\ion{O}{3}] $\lambda$4959}
\def\oiiipair{[\ion{O}{3}] $\lambda \lambda$4959,5007}
\def\taulya{\tau_{Ly\alpha}}
\def\taubar{\bar{\tau}_{Ly\alpha}}
\def\llya{L_{Ly\alpha}}
\def\ldlya{{\cal L}_{Ly\alpha}}
\def\nbar{\bar{n}}
\def\Msun{M_\odot}
\def\sqamin{\Box'}

\def\keckwin{LALA~J142441.20+353405.1}
\def\gemwin{LALA~J142442.24+353400.2}
\def\kecklose{LALA~J142544.41+353327.7}
\def\gemlose{LALA~J142610.55+354557.6}
\def\l43{L_{43}}
\def\ls{{\cal L}_{sym}}

\title{A Luminous Lyman-$\alpha$ Emitting Galaxy at Redshift z=6.535:
Discovery and Spectroscopic Confirmation%
\footnote{
The data presented in this paper were obtained at the Kitt Peak National
Observatory, the Gemini Observatory, and the W.~M. Keck Observatory.
Kitt Peak National Observatory, National Optical Astronomy Observatory,
is operated by the Association of Universities for Research in Astronomy,
Inc. (AURA) under cooperative agreement with the National Science Foundation
(NSF).
The Gemini Observatory is operated by AURA under a cooperative 
agreement with the NSF on behalf of the
Gemini partnership: the National Science Foundation (United States),
the Particle Physics and Astronomy Research Council (United Kingdom),
the National Research Council (Canada), CONICYT (Chile), the
Australian Research Council (Australia), CNPq (Brazil) and CONICET
(Argentina). 
The W.~M. Keck Observatory is operated as a scientific partnership
among the California Institute of Technology, the University of
California and the National Aeronautics and Space Administration.  The
Keck Observatory was made possible by the generous financial support
of the W.M. Keck Foundation.}}

\author{
James E. Rhoads \altaffilmark{2,3},
Chun Xu \altaffilmark{2},
Steve Dawson \altaffilmark{6},
Arjun Dey \altaffilmark{4},
Sangeeta Malhotra \altaffilmark{2,3},
JunXian Wang \altaffilmark{5},
Buell T. Jannuzi \altaffilmark{4},
Hyron Spinrad \altaffilmark{6},
Daniel Stern \altaffilmark{7}
}

\begin{abstract}
We present a redshift $z=6.535$ galaxy discovered by its \lya\
emission in a 9180\AA\ narrowband image from the Large Area Lyman
Alpha (LALA) survey.  The \lya\ line luminosity ($1.1\times 10^{43}
\ergsec$) is among the largest known for star forming galaxies at $z\approx
6.5$.  The line shows the distinct asymmetry that is characteristic of
high-redshift \lya.
The $2\sigma$ lower bound on the observer-frame equivalent width is
$>530$\AA.  This is hard to reconcile with a neutral intergalactic
medium unless the \lya\ line is intrinsically strong {\it and\/} is
emitted from its host galaxy with an intrinsic Doppler shift of
several hundred $\kms$.  If the IGM is ionized, it
corresponds to a rest frame equivalent width 
$> 40$\AA\ after correcting for \lya\ forest absorption.
We also present complete spectroscopic followup of the remaining
candidates with line flux  $> 2\times 10^{-17} \ergcm2s$ in our
$1200\sqamin$ narrowband image.  These include another galaxy
with a strong emission line at $9136$\AA\ and no detected continuum flux,
which however is most likely an \oiii\ source at $z=0.824$ based
on a weak detection of the \oiiib\ line.
\end{abstract}

\altaffiltext{2}{Space Telescope Science Institute, 3700 San Martin Drive, 
 Baltimore, MD 21218; email: rhoads, san, chunxu @stsci.edu} 
\altaffiltext{3}{Visiting Astronomer, Kitt Peak National Observatory.}
\altaffiltext{4}{National Optical Astronomy Observatory, 950 N. Cherry Ave., 
 Tucson, AZ 85719; email: adey, bjannuzi @noao.edu}
\altaffiltext{5}{Johns Hopkins University, Charles and 34th Street, 
 Bloomberg Center, Baltimore, MD 21218}
\altaffiltext{6}{University of California, Berkeley, CA;
 email: sdawson, spinrad @astro.berkeley.edu }
\altaffiltext{7}{Jet Propulsion Laboratory, California
Institute of Technology, Mail Stop 169-506, Pasadena, CA 91109;  email:
stern@zwolfkinder.jpl.nasa.gov}

\keywords{galaxies: high-redshift --- galaxies: formation --- 
galaxies: evolution --- cosmology: observations --- early universe}

\section{Introduction}
Observational study of the redshift range $6 \la z \la 7$ is crucial
for understanding the reionization of intergalactic hydrogen, which
was the most recent major phase transition for most of the baryonic
matter in the universe.  While polarization of the microwave
background measured by the Wilkinson Microwave Anisotropy Probe (WMAP)
satellite indicates that substantial
ionization had begun as early as $z \sim 15$ (Spergel et al. 2003), 
both the opaque
Gunn-Peterson troughs observed in the spectra of $z \ga 6.3$ quasars
(Becker et al. 2001; Fan et al. 2002)
and the high temperature of the intergalactic medium at $z \sim 4$
(Theuns et al. 2002; Hui \& Haiman 2003)
imply that a large part of the reionization was relatively recent,
with substantial neutral gas lasting up to $z \sim 6$.

\lya\ emission has proven the most effective tool so far for
identifying star forming galaxies at redshifts $z>6$; indeed, all but
two of the $\sim 6$ galaxies that have been spectroscopically confirmed
at $z>6$ were found through
their \lya\ line emission in either narrowband or spectroscopic
searches (Hu et al 2002 [H02]; Kodaira et al 2003 [K03]; Cuby et al 2003;
this work; Kneib et al 2004), and at least one of the continuum-selected
sources is also a \lya\ emitter (Cuby et al 2003).

We present the extension of the Large Area Lyman Alpha
(LALA) survey to the $z\approx 6.5$ window.  We have obtained spectra
for each of the three $z\approx 6.5$ candidates that pass all
photometric selection criteria.  Two of these show strong emission
lines in the narrowband filter bandpass.
In one case, the line is identified as \lya\ at $z\approx 6.535$ based
on its asymmetric profile and the absence of other detected lines down
to faint flux levels.  The second object is identified as an
\oiii\ source at $z\approx 0.824$ based on a probable ($4\sigma$) detection
of the \oiiib\ line.

\section{Imaging Observations and Analysis}
The LALA survey's $z\approx 6.5$ search is based on a deep 9180\AA\
narrowband image of our Bo\"{o}tes field, which is $36' \times 36'$ 
with center at 14:25:57 +35:32 (J2000.0).  The image was obtained using
the CCD Mosaic-1 Camera at the 4m Mayall Telescope of the Kitt Peak National
Observatory together with a custom narrowband filter, whose
central wavelength of $\lambda_c \approx 9182$\AA\ and 
full width at half maximum (FWHM) of $84$\AA\ place it in
a gap between the night sky airglow lines.

Narrowband imaging data were obtained on UT 2001 June 15--17 and
24--25 and 2002 April 5 and 18--19.  The total exposure
time was 28 hours.

Data reduction followed the same procedures used in the $z=4.5$ and
$z=5.7$ LALA images (Rhoads et al. 2000; Rhoads \& Malhotra 2001).  To
summarize, we remove electronic crosstalk between Mosaic chip pairs 
sharing readout electronics; subtract 
overscan and bias frame corrections; and flatfield with
dome flats.  Next, we subtract a pupil ghost image caused by internal
reflections in the KPNO 4m corrector.  We remove residual,
large scale imperfections in the
sky flatness using a smoothed supersky flat derived from the science
data, followed by subtraction of a polynomial surface fit to the sky
flux.  A fringe frame was constructed from the data, fitted to the
blank sky regions of each exposure, and subtracted.
Cosmic rays were rejected in each exposure using the algorithm
of Rhoads (2000).

The world coordinate systems of individual frames were adjusted to the
USNO-A2 star catalog (Monet et al 1998).
Satellite trails were flagged by hand for
exclusion from the final stacks.  The eight-chip Mosaic-1 images were
then mapped onto a rectified, common coordinate grid using the
``mscimage'' task from the IRAF MSCRED package (Valdes \& Tody 1998;
Valdes 1998).  The seeing, throughput, and sky brightness in each
exposure were measured and used to compute a set of weights for image
stacking using the ``ATTWEIGHT'' algorithm (Fischer \& Kochanski
1994). Finally, the exposures were stacked, with an additional sigma
clipping applied to reject discrepant data not previously flagged.
The majority of the weight comes from the final two nights'
imaging data, and only $24\%$ of the total weight is from the 2001
observing season.

The final stack has seeing $1.02''$ and a $5\sigma$ sensitivity
limit of $0.7 \microJy$ (measured within a 9 pixel [$2.32''$] diameter
aperture),  corresponding to $2.0 \times 10^{-17} \ergcm2s$ for
a pure line source at these wavelengths or to an AB magnitude
of $24.3$ for a pure continuum source.
The magnitude zero point was determined by comparison with a
deep $z'$ filter image obtained earlier (see Rhoads \& Malhotra 2001),
which was in turn calibrated to $z'$ filter standard stars from
SDSS standard star lists.

\section{Candidate Selection}
\lya\ galaxy candidates were selected following the criteria
used successfully in lower redshift LALA searches (Rhoads \& Malhotra
2001; Dawson et al. 2004).  In summary, these include: 1) Significance
of the narrowband detection $>5 \sigma$; 2) A narrowband excess of at
least $0.75$ magnitude, so that $\ge 50\%$ of the narrowband flux
comes from an emission line; 3) Significance of the narrowband excess
$> 4 \sigma$; and 4) No detection in filters blueward of the expected
Lyman break location at the $> 2\sigma$ level.

To implement the blue flux ``veto'' criterion, we used a weighted
sum of six NOAO Deep Wide-Field Survey (NDWFS; Jannuzi \& Dey 1999)
and LALA survey filters
blueward of $6850$\AA, where the Lyman break would fall for galaxies
whose \lya\ line falls within the $9180$\AA\ narrowband.  The weights
were chosen to achieve optimal depth for objects of approximately
median color.  Most of the weight in this combined ``veto image'' comes
from the NDWFS B$_W$ filter, with additional contributions from NDWFS R
band data and the LALA survey's V band, and three of the redshifted 
H$\alpha$ narrowband filter observations.  
This stack reaches a final $2\sigma$ limiting AB magnitude of $27.1$ 
at an effective central wavelength near $5000$\AA.
Formally, some flux may be expected in the R band filter for $z\approx 6.5$
sources, but in practice, intergalactic hydrogen absorption attenuates this
flux so severely that it is quite safe to include R band data 
in the veto stack as we have done.  

Photometry for candidate selection was done using SExtractor (Bertin
\& Arnouts 1996).
All color tests were applied using a 9 pixel diameter aperture,
corresponding to $2.32''$, and using SExtractor's two-image
mode to ensure photometry from the same regions in all filters.

In addition to high redshift \lya\ sources, three types of
contaminants may enter the sample.  The first is noise spikes in the
narrowband image.  Assuming Gaussian statistics, we should expect
$\sim 1$ noise peak above our $5\sigma$ detection threshold among the
$\sim 10^{6.5}$ independent resolution elements in the narrowband
image, and more may be found if the noise properties of the image are
not perfectly Gaussian. 
The second is time variable sources (either transient or moving
objects).  These may mimic emission line objects in our catalog
because different filters were observed at widely varying times.  For
typical LALA survey depths, the expected rate of high redshift
supernovae will be $\sim 1$ per filter per field (Riess, personal
communication; see also Strolger et al 2004).
The third is low-redshift emission line galaxies of extremely high
equivalent width.  The expected numbers of such foreground objects are
difficult to estimate from either theory or present data--- the LALA data
themselves are likely to considerably refine the statistics of such
contaminating sources.

These criteria yielded a total of three high quality candidate
$z\approx 6.5$ \lya\ galaxies.  Formally, six objects passed
selection using an old veto filter stack from 2001, but two of these
were ruled out completely and a third strongly disfavored by detections
in a new NDWFS R band image from 10 April 2002.  One of these excluded
candidates is almost certainly a supernova at moderate redshift; such an event 
would not be surprising given the area and depth of our survey.
Properties of the four best candidates are summarized in table~\ref{cantab}.

\clearpage

\begin{deluxetable}{rlllllll}
\tabletypesize{\footnotesize}
\tablecolumns{8}
\tablewidth{0pc}
\tablecaption{Photometry}
\tablehead{
\colhead{ID} & \colhead{918nm flux} &\colhead{$z'$ flux} & 
  \colhead{$I$ flux} & \colhead{$R$ flux} & \colhead{Photometric} &
  \colhead{W$_\lambda^{\rm rest} (2\sigma)$} & Nature\\
 & \colhead{($\microJy$)} & \colhead{($\microJy$)} & \colhead{($\microJy$)}
 & \colhead{($\microJy$)} & \colhead{line flux\tablenotemark{a}}
 & \colhead{limit (\AA)} & }
\startdata
LALA J142441.20 & $0.763$ & $-0.214$ & $-0.0145$ &  $-0.0372$ &  $4$ to $6.6$
\tablenotemark{b} &  $>620$ & $z=0.824$ \\
 +353405.1 & $ \pm 0.139$ & $\pm 0.111$ & $ \pm 0.045$ &  $\pm 0.029$ &  &  &  \\ 
LALA J142442.24 & $0.767$ & $-0.081$  & $-0.064$  &  $-0.061$ & $2.26$ &  $>70$ & $z=6.535$ \\ 
+353400.2 & $\pm 0.137$ & $\pm 0.113$ & $\pm 0.045$ & $\pm 0.028$ & $\pm 0.40$ &   &  \\
LALA J142544.41 & $0.706$ & $-0.051$    & $-0.073$   &  $-0.021$    & $2.08$ &  & Noise spike\\
  +353327.7 & $ \pm 0.138$& $\pm 0.106$ & $\pm 0.044$ & $\pm 0.029$ & $\pm 0.41$ &  &\\
LALA J142610.55 & $0.797$  & $-0.137$   & $ 0.011$    &  $0.057$   &   $2.35$ &  & Unknown\\ 
+354557.6    & $\pm 0.121$ &$\pm 0.110$ & $\pm 0.046$ & $\pm 0.030$ & $\pm 0.36$ &  & \\
\enddata
\tablenotetext{a}{Units of $10^{-17} \ergcm2s$.}
\tablenotetext{b}{The line for \keckwin\ falls on the edge of the 
narrowband filter, where throughput is a rapid function of wavelength.
The dominant uncertainty in the photometric line flux is the 
resulting throughput correction at $9136$\AA, and we quote a range
of flux values based on the plausible range of this throughput correction.
See text.}
\tablecomments{Photometric properties of the candidate $z\approx 6.5$
LALA sources. The $2\sigma$ lower bounds on equivalent width are derived
from the narrow- and broad-band photometry for \gemwin,
and from DEIMOS spectra for \keckwin, and are corrected to the
spectroscopically determined redshift. Tabulated values
are not corrected for IGM absorption.
\label{cantab} }
\end{deluxetable}

\clearpage

\section{Keck and Gemini Spectra}
Two of the z=6.5 candidates (\keckwin\ and \kecklose) were
observed using the DEIMOS spectrograph (Faber et al. 2002) on the Keck II
telescope of the W. M. Keck Observatory on the night of U.T. 2003 May 01.
Observations of \keckwin\ were also obtained on U.T. 2003 Apr 01.
All observations were made using the 600 l/mm ($\lambda_b = 8500$\AA)
grating through slitmasks with slit widths of 1.0$\arcsec$.  This setup
gives a dispersion of $\approx 0.63{\rm \AA}/\hbox{pixel}$ and a
spatial plate scale of $0.1185\arcsec/\hbox{pixel}$. The total
exposure times were 2.5 hours for \keckwin\ (1 hour on Apr~01 and 1.5
on May~01) and 2.0 hours for \kecklose.
These were broken into individual exposures of 1800s, with
2.5 \arcsec\ spatial offsets between exposures to facilitate the
removal of fringing at long wavelengths. The nights were clear and the
seeing was typically 0.5-0.8$\arcsec$. The data were reduced using the U.C.
Berkeley pipeline reduction software (Davis et al. 2004) adapted from
programs developed for the SDSS (Burles and Schlegel 2004), and further
adapted to our observing mode. The spectra were extracted and analyzed
using IRAF (Tody 1993).

The candidates \keckwin\ and \gemwin\ were observed on U.T. 2003 May 8
and June 29 using the GMOS spectrograph on Gemini North (Hook et
al 2003) in Nod-and-Shuffle mode (Glazebrook \& Bland-Hawthorn 2001).  
The observations were made using the R400+G5305 grating
(400 lines/mm, central wavelengths at 756 - 768 nm) through slitmasks
with slit widths of 1.0$\arcsec$. On each night, the total exposure
time was broken into 5 individual exposures of 900s per nod position.
Subsequent exposures were offset by 0.2 $\arcsec$
in the spatial direction and 3 nm in  wavelength, in
order to remove the instrumental features (such as the
horizontal stripes in the individual exposures) and to cover the chip
gaps. Since the two objects appear on both the target and sky
exposures (i.e., on both shuffle A and shuffle B), the effective
exposure time for the two sources are 9000s or 2.5 hours per night, 
for a grand total of 5 hours' on-source integration.

Additionally, a second GMOS slit mask was observed for a total of 3.5
hours total integration time on UT 2003 July 02 (0.5 hour) and July 03
(3.0 hours), covering the (lower grade) candidate \gemlose.

The data were reduced using the Gemini GMOS packages (v1.4) within IRAF
through the standard procedures (http://www.gemini.edu/). The wavelength
calibrations were performed
based on the CuAr lamp spectra and were double checked against the
night sky lines.  
The uncertainties in the extracted spectra were obtained empirically
by measuring the RMS flux among all spatially distinct pixels at each
wavelength in the rectified 2-D spectra.

Additional targets were of course included on all slit masks,
including lower redshift emission line galaxies, continuum-selected
Lyman break galaxies and intermediate redshift elliptical galaxies,
and X-ray sources from the $172$ ksec LALA Bo\"{o}tes field Chandra
ACIS observation (Wang et al. 2003; Malhotra et al. 2003).  Results for
these other targets will be presented elsewhere.

\section{Spectroscopic Results}
Two of our candidates, \gemwin\ and \keckwin, yielded emission lines
within the bandpass of the narrowband filter.  Of these, one\\ (\gemwin)
is confirmed as a \lya\ galaxy, while the other\\ (\keckwin) is
identified as \oiii.

\begin{figure}
\epsscale{0.99}
\plotone{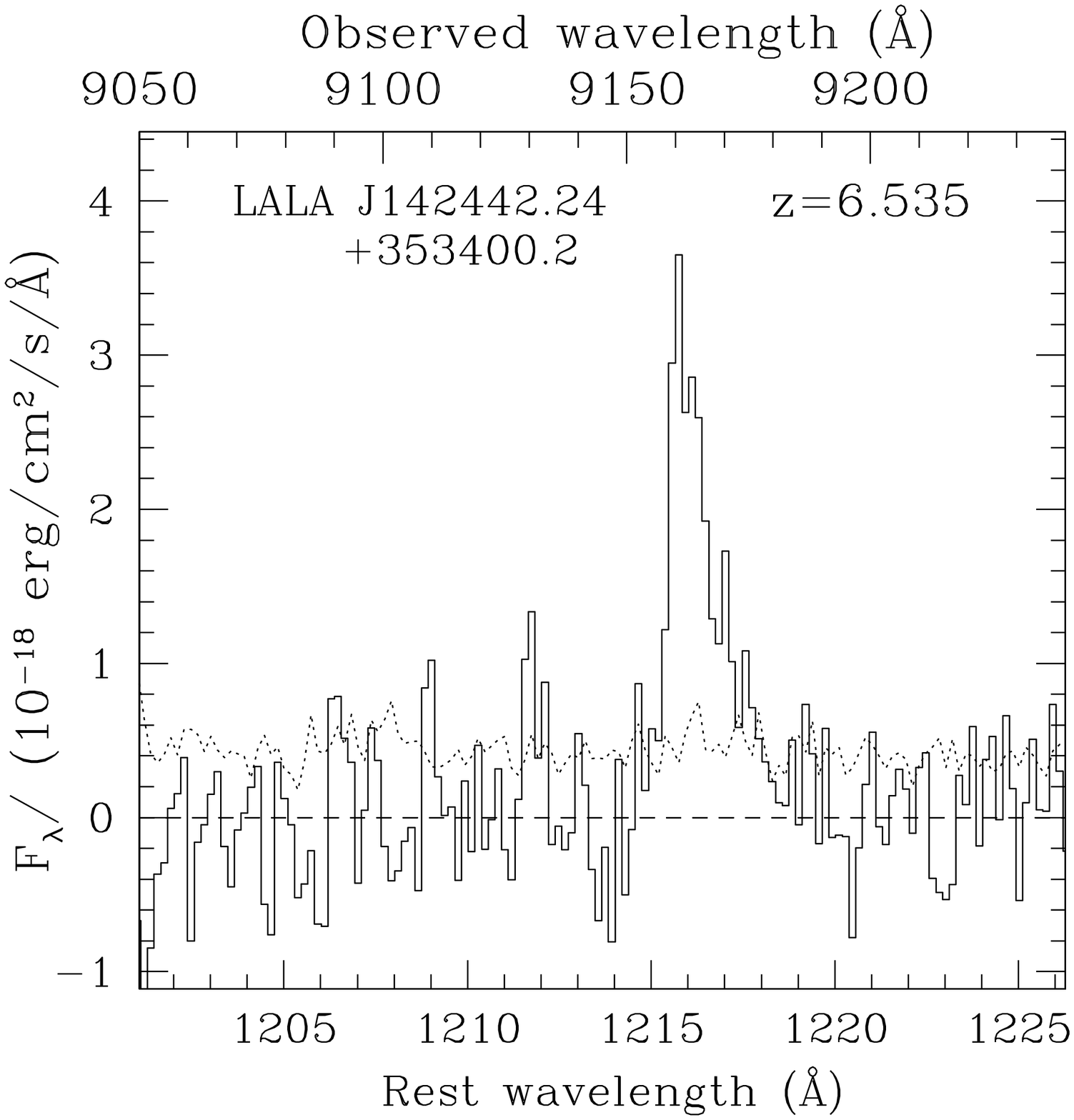}
\caption{Spectrum of \gemwin, obtained with the GMOS spectrograph
on Gemini North.  The solid histogram shows the measured flux;
the dotted line shows the one-sigma flux uncertainties.
The asymmetry of the line is clearly visible.
There is no emission at 9072\AA, where the \oiiib\ line would be
expected if the 9160\AA\ feature were \oiii.
The conversion from counts to physical units was determined
by requiring the integrated line flux to match that determined
from our narrowband photometry.
\label{gwin}}
\end{figure}

\begin{figure}
\epsscale{0.99}
\plotone{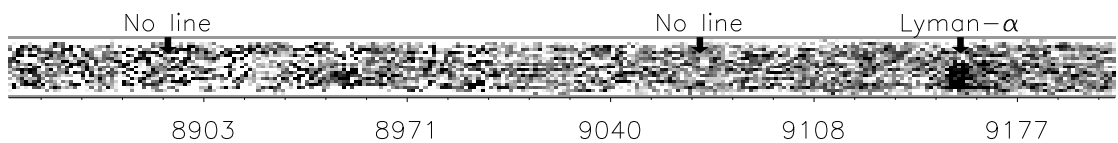}
\caption{2 dimensional GMOS spectrum of \gemwin. 
Night sky lines are well subtracted here, but their locations 
are plainly visible as bands of enhanced
photon noise.  Arrows mark the locations where the H$\beta$ and \oiiib\ lines
would be expected if the primary line at $9160$\AA\ were \oiii.
\label{gwin2d}}
\end{figure}

\paragraph{\gemwin:} \label{gwin_obs}
\gemwin\ was confirmed by Gemini-N + GMOS on 2003 May~8 and 2003
June~29.  No DEIMOS data were obtained for this object.  There
is a single isolated emission line at $9160$\AA, corresponding to
$z=6.535$ (see figures~\ref{gwin}, \ref{gwin2d}).  No other lines
are seen in the observed wavelength range, $6098{\rm \AA} \le \lambda
\le 10484{\rm \AA}$.

There is no evidence for continuum emission in either our
images or GMOS spectrum.  The observer-frame equivalent width of
the line is $> 530$\AA\ at the $2\sigma$ level.
We obtained this limit our LALA survey narrow- and broad-band photometry 
as follows. 
First, let $t_{\lambda,b}$ and $t_{\lambda,n}$ be the system
throughput in the broad and narrowband filters as a function of
wavelength, which we base on the filter transmission curves and
quantum efficiency of the Mosaic camera.  Then given the line
wavelength $\lambda_\ell$, we define $w_b \equiv (\int t_{\lambda,b}
d\lambda) / t_{\lambda,b}(\lambda_\ell)$ and $w_n \equiv (\int
t_{\lambda,n} d\lambda) / t_{\lambda,n}(\lambda_\ell)$.  For a narrow
line atop a flat (i.e., constant $f_\lambda$) continuum, one can show 
that $\eqw =
w_b w_n (f_{\lambda,n} - f_{\lambda,b}) / (f_{\lambda,b} w_b -
f_{\lambda,n} w_n)$, where $f_{\lambda,b}$ and $f_{\lambda,n}$ are the
flux densities measured in the broad and narrow band filters.  
We compute a $2\sigma$ limit on $\eqw$ by inserting $f_{\lambda,b} 
\rightarrow f_{\lambda,z,obs} + 2 d f_{\lambda,z}$ into the above expression.
We ignore the uncertainty in the narrowband flux $f_{\lambda,n}$,
because it is too small to much affect the $\eqw$ measurement.
For \lya\ at $z=6.535$, the rest frame equivalent width would
be $>70$\AA ($2\sigma$) if we do not correct for attenuation
by the intergalactic medium (IGM).  The IGM is expected to attenuate
the $z'$ filter flux by a factor of $\approx 0.36$ at $z=6.535$ (Madau 1995).
The correction to the line flux depends on 
the IGM neutral content, IGM velocity structure, and intrinsic line profile. 
For a symmetric line centered on a galaxy's systemic velocity, the
line flux would be reduced by a factor of $0.51$ in a predominantly 
ionized IGM (i.e. all of the red side flux but only $\sim 2\%$ of the 
blue side flux would be transmitted).  The IGM-corrected rest equivalent
width would then be $>40$\AA\ ($2\sigma$), which we regard as a conservative
lower bound.  A largely neutral IGM or
a line profile not symmetric about the systemic velocity would change
this result appreciably (see section~\ref{reion}).

The line width in
this source could in principle be consistent with \oiipair.  However,
this would imply a rest frame equivalent width $>215$\AA\ ($2\sigma$),
which is outside the usual range for this line: two large
continuum-selected samples (Cowie et al. 1996, Hogg et al. 1998) and one
H$\alpha$ line selected sample (Gallego et al. 1996) found no \oiipair\
sources with $\eqw > 140$\AA\ and a very small minority with $100 <
\eqw < 140$\AA.  Additionally, star forming galaxies usually show blue
continuum emission, so that our B$_w$ and R filter nondetections again
argue against \oiipair\ models.

\clearpage

The \gemwin\ emission line has a measured width of $13$\AA\ FWHM.
This substantially exceeds the instrumental resolution (estimated at
6\AA\ based on the GMOS spectrum of \keckwin), and thus allows
quantitative asymmetry measurements.  We apply the line asymmetry
statistics presented in Rhoads et al (2003).  First among these is
$\ls$, the likelihood of obtaining the observed data under the best
fitting symmetric line model.  We find $\ls \approx 0.02$ for
\gemwin, supporting the visual impression of asymmetry.

To go further and {\it quantify\/} this asymmetry, 
we have further developed the asymmetry statistics $a_\lambda$ and
$a_f$ that were presented in Rhoads et al. (2003).  These statistics
depend on determination of the wavelengths $\lambda_p$ where the
line flux peaks and $\lambda_{10,b}$, $\lambda_{10,r}$ where the
flux drops to $10\%$ of its peak value.  To mitigate the effects of
noise in the spectrum, all three wavelengths can be measured after
a light smoothing.  The asymmetry statistics
are then defined as $a_\lambda = (\lambda_{10,r}-\lambda_p)/
(\lambda_p-\lambda_{10,b})$ and $a_f = \int_{\lambda_p}^{\lambda_{10,r}}
f_\lambda d\lambda / \int_{\lambda_{10,b}}^{\lambda_p} f_\lambda d\lambda$.
To determine the appropriate smoothing, we simulated observations
of a line having a truncated gaussian profile (described below) and
the (wavelength-dependent) noise level achieved in our GMOS data.
We find that the asymmetry of the line is most reliably recovered
using a Gaussian smoothing near 3 pixels (FWHM).  We also experimented
with the threshold level for the $\lambda_{b}$ and $\lambda_r$
measurements, but found no clear benefit to raising this threshold.
Using 3~pixel smoothing, we find $a_\lambda = 1.78 \pm 0.71$ and
$a_f = 1.54 \pm 0.53$, where the error bars are drawn from simulated
spectra and defined as a ``percentile-based sigma,'' i.e., 
$84\%$ of simulated spectra lie within $\pm \delta a$ of the median
simulated $a$ statistic.  

The \gemwin\ line can be well fit with a
truncated gaussian model of the form suggested by Hu et al. (2004): The
profile is taken to be a Gaussian of width $\sigma_1$ on the red side
of a peak wavelength $\lambda_0$ and a step function to zero intensity
blueward of the peak, but then convolved with a second Gaussian of
width $\sigma_2$ to reflect instrumental resolution effects.  In
\gemwin, we obtain $\chi^2 / (\hbox{degree of freedom}) = 0.93$
using $\sigma_1 = 8$\AA, $\sigma_2 = 1.2$\AA, and $\lambda_0 =
9158$\AA.  If we fix $\sigma_2 = 2.5$\AA\ (the value
obtained by assuming the emission line in \keckwin\ is unresolved
by GMOS), the best fit deteriorates slightly to
$\chi^2 / (\hbox{degree of freedom}) =1.17$.

Given that we see no other significant lines in the spectrum of
\gemwin, we can rule out \oiii\ and H$\alpha$ interpretations quite
strongly.  The remaining possibility, besides \lya, is \oiipair.
However, the line ratio for \oiipair\ sources under usual
astrophysical conditions is $f_{3729} / f_{3726} \approx 1.3$,
yielding an asymmetry opposite that of \lya\ when the doublet is
blended.  We have tried modelling the data with \oiipair\ doublets,
broadened by the instrumental resolution and an internal
velocity dispersion (taken as a free parameter).
Fixing the line ratio at $1.3$ makes it impossible to achieve
$\chi^2/\hbox{d.o.f.} \la 1.9$, which is clearly worse than the truncated
Gaussian model for \lya.  If we allow the line ratio to vary
arbitrarily, we can achieve $\chi^2/\hbox{d.o.f.} = 1.0$, but
only for $f_{3729} / f_{3726} \approx 0.4$, which would require
an electron density $\ga 10^4 cm^{-3} * (T/1e4)^{-1/5}$ (e.g., Keenan
et al 1999).  Extragalactic HII regions do not approach these
densities, and their line ratios cluster in a tight range $1.2 \la
f_{3729} / f_{3726} \la 1.4$ (O'Dell \& Casta\~{n}eda 1984).
The observed range from galaxies at $z\sim 1$ in the DEEP2 survey is
similar, from 1.26 to 1.41, with a typical ratio of $\approx 1.36$ for
the high equivalent width emitters (J. Newman, personal communication).
If we fix the line ratio at 1.3 we find that simulated spectra yield 
$a_\lambda(\hbox{\ion{O}{2}}) > 1.78$ in only $6.7\%$ of simulations.
The corresponding probability for $a_f$ is slightly higher at
$12.9\%$.  Thus, the best interpretation of this line is clearly \lya,
from both its asymmetry and its equivalent width.

\paragraph{\keckwin:}
The second of these galaxies, \keckwin, was spectroscopically detected
in both our Keck + DEIMOS data (covering $5160$\AA\ to 9525\AA)
and Gemini-N + GMOS data (covering $6177$\AA to $10484$\AA).
A strong emission line is seen at a wavelength $9136$\AA.  This line
is most likely \oiii\ at a redshift $z=0.824$, based on a careful
search for other emission lines.  Unfortunately, the expected
locations of the {[\ion{O}{3}] $\lambda$4959} and H$\beta$ lines both
fall atop night sky OH emission features.  The interpretation of this
object depends critically on these features (with \lya\ being the best
interpretation if no secondary lines exist).

\clearpage

\begin{figure}
\epsscale{0.99}
\plotone{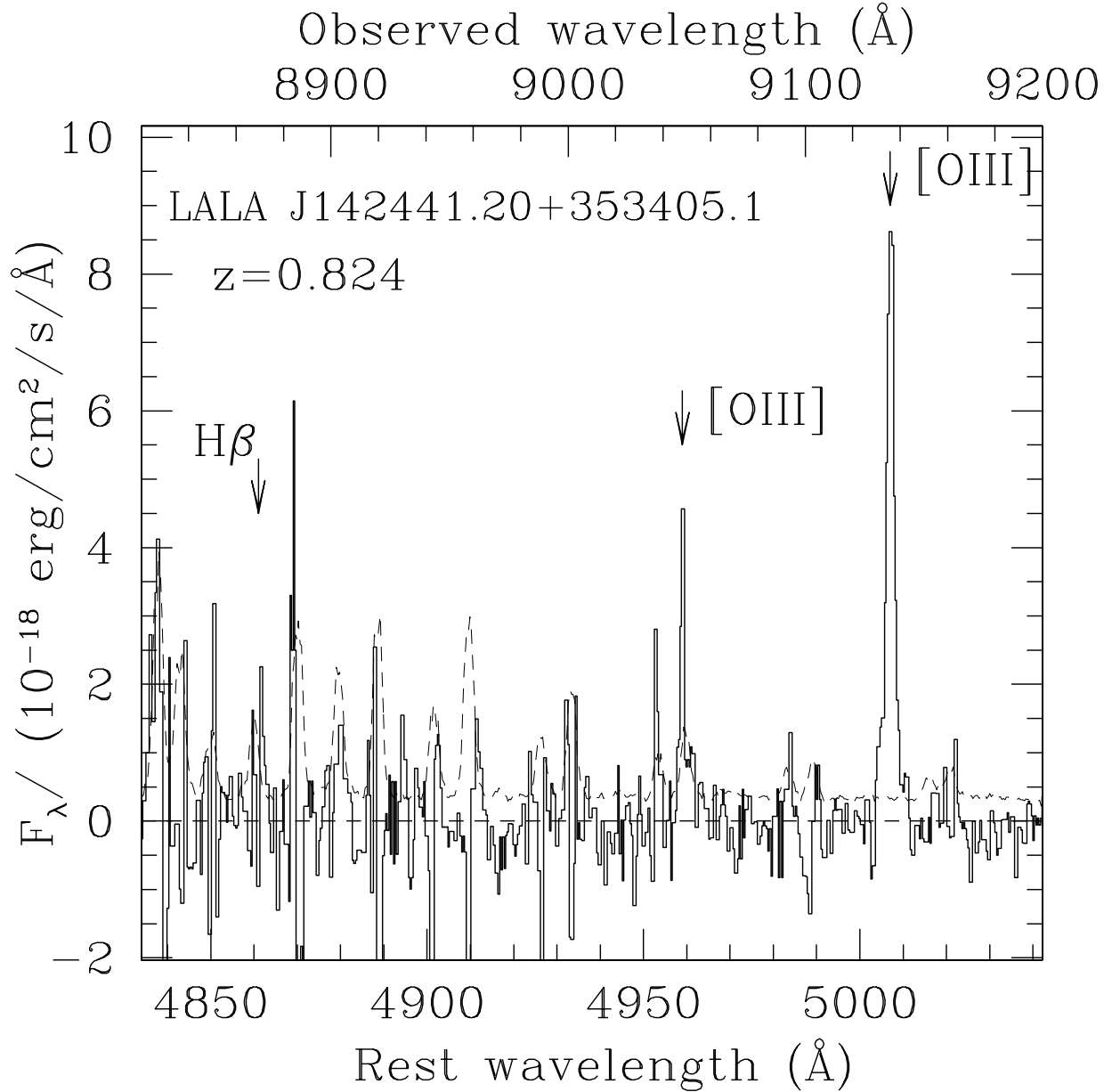}
\caption{Spectrum of \keckwin, obtained with the DEIMOS spectrograph
on Keck II.  The solid histogram shows the measured flux
after smoothing with a 3 pixel median filter to suppress residual
features from night sky lines.  The dashed line shows the $1\sigma$
flux uncertainties from photon counting errors.
We have marked the wavelengths where the
\oiiib\ and H$\beta$ lines would fall if the primary line at
$9136$\AA\ is \oiii.  
\label{kwin1d}}
\end{figure}

\begin{figure}
\epsscale{0.99}
\plotone{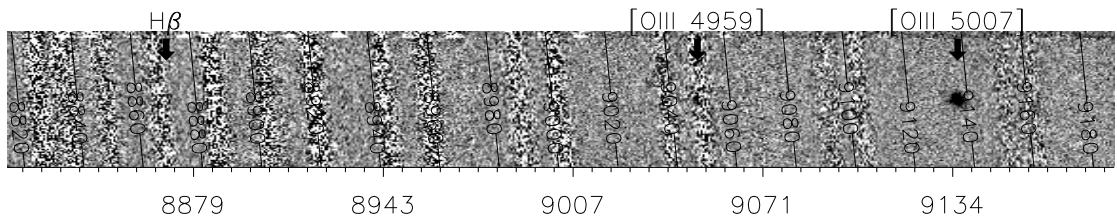}
\caption{2 dimensional DEIMOS spectrum of \keckwin.  Labeled contours
show the wavelength scale. Night sky lines are well subtracted
here, but their locations are plainly visible as bands of enhanced
photon noise.  The presumed \oiiib\ line is at 9048\AA, and is partly
visible to the blue of the 9049\AA\ night sky line. The expected
location of H$\beta$ is at 8870\AA.
\label{kwin2d}}
\end{figure}

\clearpage

We therefore fitted Gaussian line profiles
at the expected locations of both \oiiib\ and H$\beta$, using
fitting weights derived from the observed profile of
the 9136\AA\ line and from the observed noise as a function of
wavelength in sky-subtracted spectra.
This approach optimally exploits any part of the satellite
line profile that is not obscured by night sky line residuals.
We obtained a four sigma detection
of the \oiiib\ line and a line flux ratio of $0.33 \pm 0.08$
($f(4959) / f(5007)$) from our 2-D DEIMOS data.  A similar exercise
using the 1-D GMOS data gave another $4\sigma$ detection at the
expected \oiiib\ wavelength and a line ratio of $0.35 \pm 0.09$.
The consistency of these results with one another and with the
theoretically expected line ratio of $0.33$ leads us to believe
this interpretation of the data. 
Further fitting yields also a marginal H$\beta$  detection in the
DEIMOS data ($3\sigma$, line ratio $0.111 \pm 0.04$) but
nothing significant in the GMOS data.
Measurement of the \oiipair\ lines with this method is more difficult,
since the expected position of \oiipair\ depends on the spectral
trace (which we cannot measure because the continuum in \keckwin\ is
undetected).  In practice, we see a hint of the \oiipair\ doublet in both
data sets, but it is more significant in the GMOS data
($2.4\sigma$ and $6.4\sigma$ for 3726\AA\ and 3729\AA) than in
the DEIMOS data ($3.7\sigma$ and $1.0\sigma$).  
Further circumstantial support for the $z=0.824$ interpretation
comes from spectroscopic observations of a red galaxy sample in
this field, which show a strong spike in the redshift distribution
at $z\approx 0.824$ (Dey et al. 2004).

Other foreground emission line interpretations
are ruled out: An H$\alpha$ line at $9136$\AA\ should imply \oiiipair\
and H$\beta$ near 6900\AA, which is not seen, and \oiipair\ at
9136\AA\ would be easily resolved as a close doublet in our DEIMOS
data.

The flux of the $9136$\AA\ line is large but subject to systematic
uncertainties.  Flux standards from our DEIMOS run
imply a line flux of $3 \la \f17 \la 5.7$, where $\f17 \equiv
f/(10^{-17} \ergcm2s)$ is the scaled line flux.  Our LALA survey
narrowband imaging would na\"{\i}vely imply a flux $\f17 = 2.2$, but
the measured wavelength places the emission line on the wing of the
narrowband filter, where the throughput is low. Correcting by a factor
$T_{max}/T(9136{\rm \AA})$ implies a line flux $4 \la \f17 \la 6.6$,
where the lower value is based on the original filter specifications
given to the vendor and the higher value is based on filter tracings
made in May 2003 at NOAO by Heidi Schweiker and George Jacoby.  We
adopt $\f17 = 4$ as a reasonable best guess.

The line is quite narrow, with an observed FWHM of $3.55$\AA\ in our
DEIMOS data (corresponding to 5.5 DEIMOS pixels, or a velocity width
of $\approx 100 \kms$).  This width corresponds well to the typical
seeing during our observations, which means that the physical line width
in \keckwin\ could be $\ll 100 \kms$ and also that the angular size
of the line emitting region is no larger than $0.7\arcsec$.
Applying our line asymmetry test here, we find $\ls \approx 0.5$.
Because there is no demonstrable asymmetry, we did not pursue 
measurement of $a_\lambda$ and $a_f$.  A Gaussian fit to the line
yields a  $\chi^2/\hbox{d.o.f.} = 0.84$ measured from $9132$ to 9140\AA.

We also see no evidence for continuum emission.  The best equivalent
width estimate for this source comes from our DEIMOS data, which yield
a $2\sigma$ lower bound of $\eqw > 620$\AA\ in the rest frame for $z=0.824$.
While large, this is within the observed range for \oiii.  We have
previously reported a $z=0.34$ \oiii\ source with $\sim 1000$\AA\ rest 
equivalent width ($\sim 1400$\AA\ observed), discovered by its 
narrowband excess in the LALA Bo\"{o}tes field (Rhoads et al 2000).
Statistical studies of nearby HII galaxies show a range up 
to $\sim 1700$\AA\ (Raimann et al 2000).

\keckwin\ and \gemwin\ lie $13''$ apart, corresponding to a projected
separation of $\approx 80\kpc$ (physical).  This proximity appears
to be a coincidence given the differing line identifications.  If the
lines were the same transition, the velocity separation of the two
would be $\approx 800 \kms$.

\paragraph{Other candidates:}
Of the remaining two candidates, \kecklose\ was detected only in the 918nm
filter and is probably a noise spike in that filter, based on its
nondetection in Keck DEIMOS data.
Assuming Gaussian statistics, we should
expect $\sim 1$ noise peak above our $5\sigma$ detection threshold
among the $\sim 10^{6.5}$  independent resolution elements in the
narrowband image, and more may be found if the noise properties
of the image are not perfectly Gaussian.
The second spectroscopic nondetection, \gemlose, may have been a time-variable
continuum source rather than a true emission line object. 
It was targeted in a GMOS mask observed on 2003 July~02 and~03,
with a total integration time exceeding that required to detect
\keckwin\ and \gemwin, which have similar narrowband fluxes.  It appears
as a narrowband excess target when compared to images from 2001 and before,
but has a marginal R band detection in NDWFS images from UT 2002 April.  This
R band detection is offset from the 918nm source by about $0.8''$, 
and may or may not be the same source, since the faintness of the object
renders precise narrowband astrometry difficult.
Regardless of its precise nature, it appears not to be a redshift~6.5
\lya\ galaxy.

\section{Discussion}\label{discuss}

\subsection{Physical Parameters of the Sources}
Adopting a cosmology with $\Omega_m = 0.27, \Omega_\Lambda = 0.73$,
and $H_0 = 71 \kmsMpc$ (cf.\ Spergel et al. 2003) gives a luminosity
distance of $2.00\times 10^{29} \cm$ ($65 \Gpc$) for $z\approx 6.53$.
We then obtain a line  luminosity of $(1.1 \pm 0.2) \times 10^{43} \ergsec$
for \gemwin.  The other three \lya\ galaxies so far spectroscopically 
confirmed at $z\approx 6.5$ have luminosities of $1.0 \times 10^{43} \ergsec$
(SDF~J132415.7+273058; K03), $0.56 \times 10^{43} \ergsec$
(SDF~J132418.3+271455; K03), and $0.3 \times 10^{43} \ergsec$ (HCM~6A; H02).
Thus, \gemwin\ is comparable to SDF~J132415.7+273058 and may be the most 
luminous \lya\ line yet seen at this redshift.
Comparison with \lya\ luminosity function fits at $z\approx 6.5$ that
we derive in a companion paper (Malhotra et al. 2004, in preparation)
suggests that the observed line luminosity exceeds the characteristic
\lya\ luminosity $L_{Ly\alpha,\star}$ of emission line selected
galaxies at this redshift by a factor of a few.  While luminosities
substantially above $L_\star$ are often associated with strong gravitational
lensing in high redshift source populations, we consider this unlikely
for \gemwin, since any massive lensing galaxy along the line of sight
would have to have magnitude $I_{AB} \ga 25.8$ to avoid detection in
the NDWFS I band image.

We can convert the line luminosity to an estimated star formation rate
using a conversion factor of $1 \Msun/\year = 10^{42}
\ergsec$, which yields 11 $\Msun/\year$.
This conversion follows from a widely used set of
assumptions: a Kennicutt (1983) initial mass function (IMF) and 
corresponding conversion between H$\alpha$ luminosity and star
formation rate, plus the \lya\ to H$\alpha$ ratio for dust-free
case~B recombination.  These are probably not valid in detail for
high redshift \lya\ galaxies, but as we do not yet have a well
justified model, it is convenient to at least use a standard one.
Combined with our survey volume of $2.1 \times 10^5 \Mpc^3$
(comoving), we obtain a lower bound on the star formation rate density
(SFRD) of $5.2\times 10^{-5} \Msun \year^{-1} \Mpc^{-3}$ 
from \gemwin\ alone.  Of course, given that \gemwin\ is much brighter than
$L_\star$, its contribution to the star formation rate density is
only a small fraction of the total.  For comparison, K03 probe further
down the luminosity function and find a lower
limit to the SFRD of $5\times 10^{-4} \Msun \year^{-1} \Mpc^{-3}$,
while the \lya\ line found in the 
H02 survey would imply a $1\sigma$ range from $3\times 10^{-3}$ 
to $6 \times 10^{-2}$.  All of these rates are subject to incompleteness
corrections for sources falling below survey detection limits in either
\lya\ flux or equivalent width, and to additional systematic uncertainties
associated with the stellar population model assumed and with the
radiative transfer of the resonantly scattered \lya\ line.  In
particular, the model used here and elsewhere to convert between \lya\
luminosity and star formation rate cannot actually reproduce the
observed \lya\ equivalent width distribution at $z\approx 4.5$
(Malhotra \& Rhoads 2002).  Still, relative comparisons among \lya\ samples
remain valid, and show that sources like \gemwin\ are rare, luminous, 
and constitute a fraction $\la 10^{-2}$ of the global \lya\ production 
and star formation rate at $z\approx 6.5$.

For \keckwin, we have flux measurements or limits on the \oiii, \oii,
and H$\beta$ transitions.  These allow us to place general constraints
on the metallicity of this galaxy using $R_{23} \equiv
( [f(5007) + f(4959) + f(3727)] / f(H\beta))$.
We find $\log(R_{23}) \ge \log( [f(5007)+f(4959)] / f(H\beta)) = 1.1 \pm 0.12$.
The additional contribution to the oxygen flux from \oii\ is either
negligible (using the DEIMOS measurement, for which $O_{32} \equiv
[f(5007) + f(4959)]/f(3727) \approx 20$) or small (using the GMOS
measurement, for which $O_{32} \approx 2.5$).
From Kewley \& Dopita (2002), the theoretical maximum for $\log(R_{23})$
is $0.97$, achieved only for a modestly sub-solar metallicity
near $\log(O/H)+12 \approx 8.38$ and a high ionization parameter 
$q > 3\times 10^8 \cm/\sec$.  (Here the ionization parameter is defined
as the ratio of ionizing photon flux to number density of hydrogen; see
Kewley and Dopita 2002.)  Our observation is consistent with this theoretical
maximum, but not with $R_{23}$ substantially below this maximum, and we
thus conclude that \keckwin\ has $12+\log(O/H) \approx 8.4$.  For comparison,
estimates of the solar value for $12+\log(O/H)$ range from $8.9$
(McGaugh 1991) to $8.68$ (Allende Prieto, Lambert, \& Asplund 2001).

This places \keckwin\ in the lowest quartile of metallicity for 
galaxies at $0.47<z<0.92$ from the Canada-France Redshift Survey
sample (Lilly, Carollo, \& Stockton 2003), but around the midpoint of
the metallicity range for Lyman break galaxies at $z\approx 3$
(Pettini et al 2001).  The ratio $O_{32}$ also shows a closer physical
resemblance to LBGs than to continuum-bright galaxies at
$0.47<z<0.92$: Ratios $O_{32}>2$ are observed in only one of 66
galaxies reported by Lilly et al (2003) but in four of five LBGs in
the Pettini et al (2001) sample.  However, the apparent discrepancy
with other $z\sim 0.8$ galaxies is likely due to the larger characteristic
luminosity of the Lilly et al sample, which has $I_{AB} < 22.5$. 
Our I band limiting magnitude ($I_{AB} > 25.9$) corresponds to Johnson
$M_B > -16.9$.  Using a metallicity-luminosity relation derived by
Melbourne \& Salzer (2002) for local emission line selected galaxies,
we would expect $12+\log(O/H) = 8.1 \pm 0.27$ where the error bar is
the empirically determined scatter for the local sample.  Thus,
\keckwin\ actually has a slightly {\it high\/} metallicity for its
luminosity.  This is broadly consistent with the finding of Lilly et
al (2003), that most galaxies at $0.47<z<0.92$ have metallicity comparable
to local galaxies of similar luminosity.

Photoionization models show that equivalent widths of $\sim 600$\AA\
correspond to starburst ages of $3$ to $4 \Myr$ for metallicity $Z
\sim Z_\odot / 4$ (Stasi\'{n}ska \& Leitherer 1996), implying that
the emission from \keckwin\ is dominated by a very young stellar population.
Combining this age with the gas metallicity implies that this is not
the first generation of stars formed in this high-redshift dwarf galaxy.
The current star formation rate can be estimated from the emission line
fluxes.  The conversion between line flux and star formation rate is
better established for $H\beta$ (using a ratio $f(H\alpha) / f(H\beta)
= 2.8$ and the Kennicutt (1983) conversion factor) than for \oiii.
The observed $H\beta$ flux thereby implies a star formation rate is
$\sim 0.4 \Msun/\year$, though the combined uncertainty in the line flux 
and conversion factor is at least a factor of two.

\subsection{Implications for Reionization} \label{reion}
Because \lya\ photons are resonantly scattered by neutral hydrogen,
\lya\ emitting galaxies may suffer considerable attenuation of their
line flux when embedded in an intergalactic medium (IGM) with a substantial
neutral fraction.  Thus, a decrease in \lya\ galaxy counts may be
expected at redshifts substantially before the end of reionization
(Rhoads \& Malhotra 2001, Hu et al. 2002, Rhoads et al. 2003).  Typical
high redshift \lya\ galaxies have small continuum fluxes, and upper
limits can be placed on their expected Stromgren sphere radius $r_s$
by combining their observed fluxes and equivalent widths with stellar
population models (Rhoads \& Malhotra 2001).  A radius of $r_s \sim
0.5 \Mpc$ is typically inferred.  For comparison, $r_s > 1.2 \Mpc$ is
required to avoid a scattering optical depth $\tau > 1$ at the systemic
velocity due to the neutral IGM outside the Stromgren sphere.

Haiman (2002) has argued that the first reported $z\approx 6.5$ galaxy
(Hu et al. 2002) could still be embedded in a neutral IGM, because (a)
emission from the red wing of the emitted line will be less strongly
scattered than line center, reducing the effective attenuation of the
line, and (b) the observed \lya\ equivalent width in this object is
substantially below the theoretical maximum from stellar population
models, so that the observed line could in fact be substantially
attenuated.  Santos (2003) considers additional physical effects,
including gas motions and overdensity associated with the formation of
\lya\ galaxies, which generally increase the \lya\ attenuation over
that in Haiman's model.  For example, in the case of infall to a forming
galaxy, scattering by residual neutrals within the Stromgren sphere
can produce large optical depths ($\tau \sim 100$) on the blue
side of the \lya\ line, between the systemic velocity and the gas 
infall velocity.  However, Santos also suggests that the
velocity differences observed between \lya\ lines and other emission
lines in Lyman break galaxies could be intrinsic to the source, so that
the entire \lya\ line is emitted with an effective redshift of a few hundred 
$\kms$.  In this case, the effective \lya\ attenuation for $z\sim 6.5$ is a 
factor of $\sim 3$, and is not very sensitive to model parameters
besides the ionization state of the IGM.

\gemwin\ constrains reionization better than most other $z\approx 6.5$
galaxies because of its relatively large equivalent width ($>70$\AA\
rest frame, $2\sigma$, before corrections for IGM opacity; or $>40$\AA
with corrections suitable for an ionized IGM; see section~\ref{gwin_obs}).
The source SDF~J132415.7+273058 (K03) has a
similar line luminosity, a continuum detection near the upper limit
for \gemwin, and hence an equivalent width measurement quite similar
to \gemwin's $2\sigma$ limit.

In a neutral universe, these sources would require an intrinsic
equivalent width $\ga 120$\AA\ even if the \lya\ line is redshifted as
suggested by Santos, and $\ga 400$\AA\ in the absence of such a
redshift.  The former is consistent with the observed \lya\ equivalent
width distribution at $z\approx 4.5$ (Malhotra \& Rhoads 2002).  The
latter is high even for the $z\approx 4.5$ sample. Thus, the
properties of these objects are most easily understood if the universe
is mostly ionized at $z\approx 6.5$.  More robust constraints on
reionization from {\it individual} \lya\ galaxies will be difficult,
because the expected suppression of the \lya\ line depends strongly on
the velocity offset between emitted \lya\ and the galaxy's systemic
velocity, and measuring the latter would in general require
spectroscopy in the thermally dominated mid-infrared (Santos 2003).

Stronger constraints on reionization from \lya\ galaxies
will be possible using statistical samples.  Suppression of \lya\ by
the IGM will have a strong effect on the \lya\ luminosity function:
Even a factor of $\sim 3$ reduction in \lya\ flux will reduce the
observed \lya\ number counts dramatically.  The accessible portion of
the \lya\ luminosity function is very steep (see Malhotra et al. 2004),
and a factor of $3$ in luminosity threshold corresponds to a factor of
$\ga 10$ in expected \lya\ source counts at fixed sensitivity.
Moreover, the effect on the equivalent width distribution will be identical
to that on the luminosity function, which would not be generically
expected for other factors that could modify the \lya\ luminosity
function (such as evolution).

\section{Conclusions}
We have performed a narrowband search for \lya\ emitting galaxies
at $z\approx 6.5$ in the Bo\"{o}tes field of the Large Area Lyman
Alpha survey, and have obtained spectra for all of our viable
candidates.  One source, \gemwin, is confirmed as \lya\ at $z =
6.535$, based on an isolated, asymmetric emission line.
The \lya\ luminosity of this source is large, $\approx 1.1\times 10^{43}
\ergsec$.  Objects this bright in the line are rare, occurring
at a rate $\sim 1$ per $2\times 10^5 \hbox{comoving} \Mpc^3$ (this work
and K03).
It is therefore likely to correspond to a high peak in
the density distribution at redshift $z=6.535$.  The equivalent
width is also larger than most other $z\approx 6.5$ galaxies, with
a $2\sigma$ lower bound of $>40$\AA\ (rest frame, corrected for
IGM absorption).  This is difficult
to reconcile with a neutral intergalactic medium unless the
\lya\ line is  intrinsically strong {\it and\/} is emitted
from its host galaxy with an intrinsic Doppler shift of several
hundred $\kms$.

\acknowledgements 
We thank Paul Groot and collaborators for their help in obtaining some
of our June 2001 imaging data;
Richard Green, Jim De Veny, and Bruce Bohannon
for their support of the LALA survey;
Heidi Schweiker and George Jacoby for help with filter curve measurements;
and an anonymous referee, for a prompt and helpful report.
This work made use of images provided by the NOAO Deep Wide-Field
Survey (Jannuzi and Dey 1999), which is supported by the National
Optical Astronomy Observatory (NOAO).  We thank Melissa Miller,
Alyson Ford, Michael J. I. Brown, and the rest of the NDFWS team for
their work on the NDFWS data.  NOAO is operated by AURA,
Inc., under a cooperative agreement with the National Science
Foundation.
STScI is operated by the Association of Universities for Research in
Astronomy, Inc., under NASA contract NAS5-26555.
%
The analysis pipeline used to reduce the DEIMOS data was developed at UC
Berkeley with support from NSF grant AST-0071048.
The work of S.~D. was supported by IGPP-LLNL University Collaborative
Research Program grant 02-AP-015 and was performed under the auspices of
the Department of Energy, National Nuclear Security Administration, by the
University of California, Lawrence Livermore National Laboratory, under
contract W-7405-Eng-48.
H.~S. thanks the National Science Foundation for support under
NSF grant AST-0097163.
The work of D.~S. was carried out at Jet Propulsion Laboratory, California 
Institute of Technology, under a contract with NASA.
The authors wish to recognize and acknowledge the very significant
cultural role and reverence that the summit of Mauna Kea has always had
within the indigenous Hawaiian community.  We are most fortunate to have
the opportunity to conduct observations from this mountain.

\end{document}